\documentclass{llncs}
\usepackage{epsfig}
\usepackage{amssymb}
\usepackage{amsmath}
\usepackage{mathtools}
\usepackage{boxedminipage}
\usepackage{graphicx}
\usepackage[ruled,algosection,vlined,linesnumbered]{algorithm2e}

\graphicspath{{figures/}{fig/}{./}}

\SetKwBlock{Event}{on event}{end}
\SetKw{Break}{break}
\SetKw{Cont}{continue}

\SetAlFnt{\scriptsize}
\begin{document}

\title{A Protocol for the Atomic Capture of Multiple Molecules on Large Scale Platforms}

\author{Marin Bertier \and Marko Obrovac \and C\'edric Tedeschi}
\institute{IRISA / INRIA, France
\\ \textit{firstname.lastname@inria.fr}}

%\date{\mbox{}}

\maketitle

\begin{abstract}

With the rise of service-oriented computing, applications are more and more
based on coordination of autonomous services. Envisioned over largely
distributed and highly dynamic platforms, expressing this coordination calls for
alternative programming models. The chemical programming paradigm, which models
applications as chemical solutions where molecules representing digital entities
involved in the computation, react together to produce a result, has been
recently shown to provide the needed abstractions for autonomic coordination of
services.

However, the execution of such programs over large scale platforms raises
several problems hindering this paradigm to be actually leveraged. Among them,
the atomic capture of molecules participating in concurrent reactions is one of
the most significant.

In this paper, we propose a protocol for the atomic capture of these molecules
distributed and evolving over a large scale platform. As the density of possible
reactions is crucial for the liveness and efficiency of such a capture,  the
protocol proposed is made up of two sub-protocols, each of them aimed at addressing
different levels of densities of potential reactions in the solution. While the
decision to choose one or the other is local to each node participating in a
program's execution, a global coherent behaviour is obtained. Proof of liveness, as
well as intensive simulation results showing the efficiency and limited overhead
of the protocol are given.

\end{abstract}

\section{Introduction}
\label{sec:intro}

% General context

With the widespread adoption of the Service-Oriented Architecture (SOA)
paradigm, large scale computing platforms have recently undergone a new shift in
their shape and usage. Within such platforms, the basic entity is a
\emph{service}, \emph{i.e.}, an encapsulation of some computing, storage, or
sensor device, to be used by users or combined with other services. On top of
these platforms, applications now commonly compose these services dynamically,
under the shape of \emph{workflows}, \emph{i.e.},  temporal compositions of
services. To run over emerging highly distributed and dynamic platforms, without
any central authority or orchestrator, services need to be able to coordinate
themselves autonomously, in a fully-distributed fashion. In this context,
programming models need to be rethought in order to provide the right
abstractions for this coordination, while taking into account the distribution
and dynamics of the underlying platform.

%% Chemical programming

Artificial chemistries~\cite{Dittrich2001}, which are
chemically-inspired information processing models, have regained
\emph{momentum} in this context, and are now used to model this
\emph{ecosystem} of services~\cite{VZ09}. More concretely, the
chemical programming paradigm, initially developed to write highly
parallel programs, was identified to provide the right level of
abstraction for this context~\cite{BP09}. Within the basic version of
the chemical programming model~\cite{par_gamma_machine}, a program is
envisioned as a \emph{chemical solution} where molecules of data float
and react according to some \emph{reaction rules} specifying the
program, to produce new data (products of reactions).
%More formally
%speaking, the solution is a multiset of data, and the reactions are
%modelled by rewriting rules on this multiset.
At runtime, reactions
arise in an implicitly autonomous and parallel mode, and in a
non-deterministic order. When no more reactions are possible,
%the
%program is said to be \emph{inert}. In this state,
the solution
contains the result of the computation. 
%In its higher-order
%version~\cite{BFR06}, chemical programming allows to describe any
%component of a virtual system as a molecule. In particular, this
%feature offers the possibility of having rules reacting with other
%rules, \emph{i.e.}, programs modifying programs, at runtime.

%% Pb, transition to distributed systems.

While the chemical paradigm allows the easy design of coordination protocols,
running these chemical specifications over distributed platforms is still a
widely open issue. Among one of the most significant barriers to be lifted is
the atomic capture of multiple molecules satisfying a reaction. At
runtime, a molecule can potentially participate in several
concurrent reactions. However, it is allowed to participate in only one. Otherwise, the
logic of the program would be broken. This problem is exemplified in
Section~\ref{sec:prelim}.

%% More precisely defining the pb, also in regards

Let us slightly refine the problem considered in this paper. We consider a
chemical program made of a multiset of data, and of a set of rules acting
concurrently on them. Both data and rules are distributed amongst a set of nodes
on which the program runs. Each node periodically tries to fetch molecules
needed for their assigned reactions. As several molecules can satisfy the
pattern and conditions of several reactions performed concurrently by different
nodes, the same molecule can be requested by several nodes at the same time,
inevitably leading to conflicts. Mutual exclusion on the molecules is thus
mandatory.

Although our problem resembles the classic resource allocation
problem~\cite{Lamport:1978}, it differs in several aspects. Firstly,
the molecules are exchangeable to some extent. Molecules requested
must match a pattern defined in the reaction rule a node wants to
perform. In other words, we differentiate two processes which are i)
finding molecules matching a pattern (achieved by a \emph{discovery
protocol}), and ii) obtaining them to perform reactions (achieved
by a \emph{capture protocol}).

Secondly --- and following the previous point --- the platform
envisioned is at large scale, and the resources dispatched over the
nodes are dynamic: molecules are deleted when they react, and new ones
are created.
%(note that the molecules consumed and those created may
%satisfy the same pattern).
Thus, the protocol to discover
molecules should be scalable and dynamic. Likewise, the number of
resources/molecules (and possible reactions) will fluctuate over time,
influencing the design of the capture protocol. Bear in mind that once
the holder of a matching molecule is located, the scale of the network
is of less importance, since only the requester and holder of the
molecule are involved in the capture protocol.

Finally, and to sum up, our objective is to define a protocol for the atomic
capture of multiple molecules, that dynamically and efficiently adapts to the
density of potential reactions in the system.

\paragraph{Contribution.}

Our contribution is a distributed protocol mixing two sub-protocols inspired by
previous works on distributed resource allocation, and adapted to the
distributed runtime of \emph{chemical} programs. 

The first sub-protocol, referred to as the \emph{optimistic} one, assumes that
the number of molecules satisfying some reaction's pattern and condition is
high, so only few conflicts for molecules will arise, nodes being likely to be
able to grab distinct sets of molecules. While this protocol is simple, fast,
and has a limited communication overhead, it does not ensure liveness when the
number of conflicts increases. The second one, called \emph{pessimistic},
slower, and more costly in terms of communication, ensures liveness in presence
of an arbitrary number of conflicts. Switching from one protocol to the other is
achieved in a scalable, distributed fashion, based on local success histories in
grabbing molecules. A proof of liveness of our protocol is given, and its
efficiency is discussed through a set of simulation results. Note that this
work, to our knowledge, pioneers research on the distributed execution of
\emph{chemical} programs.

\paragraph{Organisation of the paper.}

The next section presents the chemical programming paradigm in more details,
highlights the need for the atomic capture and describes the system model used
throughout the paper. Section~\ref{sec:protocols} details
the sub-protocols, their coexistence, and the switch from one to the other
one. Proofs of liveness and fairness are also given for the complete
protocol. Section~\ref{sec:sim} presents the simulation results and discusses
the efficiency and overhead of the protocol. Related works
%, both in the chemical
%programming and the distributed systems areas
are presented in
Section~\ref{sec:related}. Section~\ref{sec:conclusion} concludes.

% \fixme{BEGIN TODOS}

% \begin{itemize}
% 	\item put the accent on the program itself (rule-driven execution instead of
% 	the 'blind' approach employed previously in ICPP paper)
% 	\item now the focus shifts from inertia detection to smart candidate selection
% 	--- ideal for long-running and infinite tasks (e.g. monitoring)
% 	\item concrete examples of such applications (with a bit of explication):
% 	BLAST, protein construction, chemical equations, etc\ldots
% \end{itemize}

% \fixme{END TODOS}

\section{Preliminaries}
\label{sec:prelim}

Different systems require different algorithms for performing atomic operations
varying in complexity. This section describes the programming and system models
which compose the  required conditions for the protocol proposed.

\subsection{Chemical Programming Model}
\label{sec:prelim_hocl}

The chemical model was initially proposed for a natural expression of parallel
processing, by removing artificial structuring and serialisation of programs,
focusing only on the problem logic. Following the chemical analogy, data are
molecules floating in a solution. They are consumed according to some reaction
rules, \emph{i.e.}, the program, producing new molecules, \emph{i.e.}, resulting
data. These reactions take place in an implicitly parallel and autonomous way,
until no more reactions are possible, a state referred to as
\emph{inertia}. This model was first formalised by
GAMMA~\cite{par_gamma_machine}, in which the solution is a multiset of
molecules, and reactions are rewriting rules on this multiset. A rule
\textbf{replace} $P$ \textbf{by} $M$ \textbf{if} $V$ consumes a set of molecules
$N$ satisfying the pattern $P$ and the condition $V$, and produces a set of
molecules $M$. We want to emphasise here that consumption is the only possible
change of state a molecule can be subjected to: once it has been consumed, it
vanishes from the multiset completely, meaning molecules are only created and
deleted, never updated nor recreated. For the sake of illustration, let us
consider the following chemical program made up of two rules applied on a multiset
of strings, that counts the aggregated  number of characters in words with more
than two letters:

\begin{tabbing}
  xxx \= xxx \= x \= xxxx \kill
\> $\textbf{let } count = \textbf{ replace } s::string \textbf{ by } len(s)
\textbf{ if } len(s) >= 2
\textbf{ in }$\\
\> $\textbf{let }\ aggregate = \textbf{ replace } x::int, y::int \textbf{ by } x+y \textbf{ in }$\\
\> \> $\langle$\> $"maecenas",  "ligula",  "massa",  "varius",  "a", "semper"$\\
\> \> \> $"congue", "euismod", "non", "mi"~\rangle$
\end{tabbing}

The rule named \emph{count} consumes a string if it is composed of at least two
characters, and introduces an integer representing its length into the
solution. The \emph{aggregate} rule consumes two integers to produce their
sum. By its repeated execution, this rule aggregates the sums to produce the
final number. At runtime, these rules are executed repeatedly and concurrently,
the first one producing inputs for the second one. While the result of the
computation is deterministic, the order of its execution is not. Only the mutual
exclusion of reactions by the atomic capture of the reactants is implicitly
required by the paradigm.

A possible execution is the following. Let us consider, \emph{arbitrarily}, that
the first rule is applied on the first three strings as represented above, and
on the last one. The state of the multiset is then the following: $
\langle~"varius",  "a", "semper", "congue", "euismod", "non", 8, 6, 5, 2~\rangle
$ .

Then, let us assume, still arbitrarily, that the \emph{aggregate} rule is
triggered three times on the previously introduced integers, producing their
sum. Meanwhile, concurrently, the remaining strings are scanned by the
\emph{count} rule. The multiset is then: $\langle~6, "a", 6, 6, 7, 3, 2,
21~\rangle$ . With the repeated application of the \emph{aggregate} rule, the
inertia is reached ($"a"$ satisfies neither of the two rules' conditions but
could be removed with a different rule): $\langle~"a", 51~\rangle$ .

It is important to notice that the atomic capture is a fundamental
condition. Let us simply assume that the same string is captured by
different nodes running the \emph{count} rule in parallel, then the
count for a word may appear more than once in the solution, which would
obviously lead to an incorrect result.

% higher order

In the higher-order version of the chemical programming model~\cite{BFR06} any
entity taking part in the computation is represented as a molecule (including
rules), which unleashes 
%Reaction
%rules themselves react with other rules, programs modifying programs. These
%aspects confer oin HOCL
an uncommonly high expressiveness, able to naturally deal with a
wide variety of coordination patterns encountered in large scale platforms~\cite{BP09}.
However, these works remained mostly conceptual until now.

\subsection{System Model}
\label{sec:prelim_system}

We consider a distributed system $\mathbb{DS}$ consisting of $n$ machines which
communicate by message passing. They are interconnected in such way that a
message sent from a machine can be delivered, in a finite amount of time, to any other node
in $\mathbb{DS}$. At large scale, this can be achieved by relying on P2P
systems, more specifically ones employing distributed hash table (DHT)
communication protocols~\cite{chord,pastry}. They allow us to focus on the atomic
capture of molecules without having to worry about the underlying communication.

%\subsubsection{Architecture.}

\paragraph{Data and Rules Dissemination.} In the following, we assume data and
rules have already been dispatched to nodes. Note that any DHT algorithm or
network topology may be used for this purpose. Even if the data and rules are
initially held by a single external application, it can contact a node in the
DHT and transfer it the chemical solution to be executed. The node which
received the data scatters the molecules across the overlay according to the DHT's
hash function. Molecules are routed concurrently according to DHT's routing
scheme. The dissemination of rules can follow a similar
pattern, or be broadcast into the network. The only difference is that rules
can be replicated on several nodes to satisfy an increased level of
parallelism. A more accurate discussion of the rules' distribution falls out of
the scope of this paper. In the following, we simply assume every rule of the
program is present on at least one node in the system.

% In the course of the routing process, the path of
% each molecule is traced by intermediary nodes. By passing on molecules, these
% nodes maintain a \emph{local state} (in addition to the DHT's routing table)
% containing the set of nodes to which they forwarded a molecule. After the
% dissemination of molecules, the rules to execute are spread in the system using
% the nodes' local states. Once a node receives the rules to execute, its
% execution cycle starts. A more detailed description can be found in \reftodo.

\paragraph{Discovery Protocol.}  In order for the reaction to happen, a suitable
combination of molecules has to be found. While the details of this aspect are
also abstracted out in the following, it deserves to be preliminarily
discussed. The basic \emph{lookup} mechanism offered by DHTs allows the
retrieval of an object according to its (unique) identifier. Unlike the
\emph{exact match} functionality provided by DHTs, we require nodes
%Note that in this case, the hash
%function distributes data in a random and uniform fashion. This, however, will
%not match our requirements, as nodes need
to be able to find \emph{some}
molecule satisfying a pattern (\emph{e.g.}, one \emph{integer}) and condition
(\emph{e.g.}, \emph{greater than $3$}), as stated in
Section~\ref{sec:prelim_hocl}. This  can be achieved by the support of
range queries on top of the overlay network, \emph{i.e.}, mechanisms to find
some (at least one) molecules falling within a range, provided the molecules can
be totally ordered on some (possibly complex, multi-dimensional)
criterion~\cite{SP08}. This mechanism can
be easily extended to support patterns and conditions involving several
molecules. For instance, when trying to capture two molecules ordered in some
specific ways, a \emph{rule translator} --- a unit which constructs the range
query ---, based on the given rule and the first molecule
obtained, constructs the range query to be dispatched to the DHT. If matching
molecules are found, the capture protocol will be triggered.

% pread in the
% network. More specifically, it parses and examines two parts of the rule: its
% input pattern $P$ and the reaction condition $V$. The former is consulted in
% order to determine the missing molecule's type. The latter is examined in
% conjunction with the supplied molecule to supply information about the missing
% molecule's properties to the range query. In doing so, the rule translator is
% able to construct a range query complete with information about the type of the
% molecule needed to react with the previously chosen one, together with specific
% properties it must hold.
% After its construction, the range query floods the network. The exact methods
% and algorithms are not discussed here. We assume the employed method will
% correctly deliver the query to all of the nodes concerned. 

\paragraph{Fault tolerance.} DHT systems inherently provide a fault-tolerant
communication mechanism. If nodes crash, leave or join, the
communication pattern will be preserved. On top of that, in this paper
we assume there exists a higher-level resilience mechanism which
prevents loss of molecules, such as state machine
replication~\cite{smr,atomic_dht}. Each node replicates its complete
state --- the molecules and its current actions --- across $k$
neighbouring nodes. Thus, in case of its failure, one of its
neighbours is able to assume its responsibilities and continue the
computation.

\section{Protocol}
\label{sec:protocols}

Here, the protocol in charge of the atomic capture of molecules is
discussed. The protocol can run in two modes, based on two different
sub-protocols: an \emph{optimistic} and a \emph{pessimistic} one. The former is
a simplified sub-protocol which is employed while the ratio between actual and
possible reactions is kept high. When this rate drops below a certain threshold,
the latter, pessimistic sub-protocol is activated. While being the heavier of the
two in terms of network traffic, this sub-protocol ensures the liveness of the
protocol, even when an elevated number of nodes in the system compete for the
same molecules.

%First, the architecture on top of which these protocols are run is described.
%Section~\reftodo describe(s) the protocols, after which we detail how and when
%does the system switch from one protocol to the other. At the end of the
%section, an analysis of the protocols is presented.

\subsection{Pessimistic Sub-protocol}
\label{sec:protocols_pess}

Based on the three-phase commit protocol~\cite{3PC}, this sub-protocol ensures
that at least one node wanting to execute a reaction will succeed. Molecule
fetching is done in three phases --- the \emph{query}, \emph{commitment} and
\emph{fetch} phases --- and involves at least two nodes --- the node requesting
the molecules, called requester, and the nodes holding the molecules, called
holders. Algorithms \ref{alg:pes_req} and \ref{alg:pes_hol} represent the code
run on these two entities, respectively, while Figure~\ref{fig:pessimist}
delivers the time diagram of molecule fetching.

When molecules suitable for a reaction have been found (line~\ref{alg:pes_req_comb} in
Algorithm~\ref{alg:pes_req}), the query phase begins
(line~\ref{alg:pes_req_query}). The requester sends \emph{QUERY} messages
asynchronously to all of the holders to inform them it is interested in the
molecule. Depending on their local states, each of the holders evaluates
separately the received message (lines
\ref{alg:pes_hol_msg_recv}---\ref{alg:pes_hol_end_query} in
Algorithm~\ref{alg:pes_hol}) and replies with one of the following messages:
\emph{RESP\_OK} (the requested molecule is available), \emph{RESP\_REMOVED}
(the requested molecule no longer exists) and \emph{RESP\_TAKEN} (the molecule
has already been promised to another node).
%\begin{enumerate}
%	\item \emph{RESP\_OK}: the requested molecule is available
%	\item \emph{RESP\_OK}: the requested molecule still exists and there are no
%	other nodes with higher priority requesting the same molecule at this time,
%	\item \emph{RESP\_REMOVED}: the requested molecule no longer exists,
%	\item \emph{RESP\_TAKEN}: the molecule has already been promised to another
%	node.
%	\item \emph{RESP\_TAKEN}: either the molecule has already been promised to
%	another node, or there is another requester with a higher priority.
%\end{enumerate}
Unless it received only \emph{RESP\_OK} messages, the requester aborts the fetch
and issues \emph{GIVE\_UP} messages to holders, informing them it no longer
intends to fetch their molecules (line~\ref{alg:pes_req_query_resp} in Algorithm
~\ref{alg:pes_req}).

Following the query phase is the commitment phase, when the requester tries to
secure its position by asking the guarantee from the holders it will be able to
fetch the molecules (line~\ref{alg:pes_req_commit} in
Algorithm~\ref{alg:pes_req}). It does so using \emph{COMMITMENT} messages.
Upon its receipt, each holder sorts all of the requests received during the
query phase (line~\ref{alg:pes_hol_commit} in Algorithm~\ref{alg:pes_hol})
according to the conflict resolution policy (described below). Holders reply,
once again, with \emph{RESP\_OK}, \emph{RESP\_REMOVED} or \emph{RESP\_TAKEN}
messages. A \emph{RESP\_OK} response represents a holder's commitment to deliver
its molecule in the last phase. Thus, subsequent \emph{QUERY} and
\emph{COMMITMENT} requests from other nodes will be resolved with a
\emph{RESP\_TAKEN} message. Naturally, if a requester does not receive only
\emph{RESP\_OK} responses to its \emph{COMMITMENT} requests, it aborts the fetch
with \emph{GIVE\_UP} messages.

Finally, in the fetch phase, the requester issues \emph{FETCH} messages, upon
which holders transmit it the requested molecules using \emph{RESP\_MOLECULE}
messages. From this point on, holders issue \emph{RESP\_REMOVED} messages to
nodes requesting the molecule.

\paragraph{Conflict Resolution.} Each of the holders individually decides to
which requester a molecule will be given. Since at least one requester needs to
be able to complete its combination of molecules, all holders apply the same
conflict resolution scheme (lines
\ref{alg:pes_hol_sort_begin}---\ref{alg:pes_hol_sort_end} in Algorithm
\ref{alg:pes_hol}).
%Any conflict resolution scheme based on a total ordering of
%nodes could be valid. For the sake of illustration
We here detail a dynamic and 
load-balancing based scheme: each of the messages sent by requesters contains
two fields --- the requester's id and the number of reactions it has completed
thus far. When two or more requesters are competing for the same molecule,
holders give priority to the requester with the lowest number of reactions. In
case of a dispute, the requester with a lower node identifier gets the molecule.
\begin{table}[!htpb]
\begin{minipage}[t]{0.5\linewidth}
\vspace{0pt}

\begin{algorithm}[H]
	\caption{Pessimistic Protocol --- Requester.}
	\label{alg:pes_req}
	\SetKwData{Phase}{phase}
	\SetKwData{Mol}{molecule}
	\SetKwData{Resp}{response}
	\SetKwData{RespMol}{resp\_mol}
	\SetKwData{Comb}{combination}
	\SetKwData{RArgs}{reaction\_args}
	\SetKwData{RespOk}{RESP\_OK}
	\SetKwFunction{Query}{QUERY}
	\SetKwFunction{GiveUp}{GIVE\_UP}
	\SetKwFunction{Commit}{COMMITMENT}
	\SetKwFunction{Fetch}{FETCH}
	\SetKwFunction{PhQuery}{QueryPhase}
	\SetKwFunction{PhQueryResp}{QueryPhaseResp}
	\SetKwFunction{PhCommit}{CommitmentPhase}
	\SetKwFunction{PhCommitResp}{CommitmentPhaseResp}
	\SetKwFunction{PhFetch}{FetchPhase}
	\SetKwFunction{PhFetchResp}{FetchPhaseResp}
	\SetKwFunction{Abandon}{Abandon}
	\SetKwFunction{Reaction}{Reaction}
	\Event(combination found){ \label{alg:pes_req_comb}
		\PhQuery{\Comb}\;
	}
	\Event(response received){
		\If{$\Phase = query$}{
			\PhQueryResp{\RespMol}\;
		}
		\ElseIf{$\Phase = commitment$}{
			\PhCommitResp{\RespMol}\;
		}
		\ElseIf{$\Phase = fetch$}{
			\PhFetchResp{\RespMol}\;
		}
	}
	\Begin(\PhQuery{\Comb}){ \label{alg:pes_req_query}
		$\Phase \Leftarrow query$\;
		\ForEach{\Mol in \Comb}{
			dispatch \Query{\Mol}\;
		}
	}
	\Begin(\PhQueryResp{\RespMol}){ \label{alg:pes_req_query_resp}
		\If{$\RespMol \neq \RespOk$}{
			\Abandon{\Comb}\;
		}
		\ElseIf{all responses have arrived}{
			\PhCommit{\Comb}\;
		}
	}
	\Begin(\PhCommit{\Comb}){ \label{alg:pes_req_commit}
		$\Phase \Leftarrow commitment$\;
		\ForEach{\Mol in \Comb}{
			dispatch \Commit{\Mol}\;
		}
	}
	\Begin(\PhCommitResp{\RespMol}){ \label{alg:pes_req_commit_resp}
		\If{$\RespMol \neq \RespOk$}{
			\Abandon{\Comb}\;
		}
		\ElseIf{all responses have arrived}{
			\PhFetch{\Comb}\;
		}
	}
	\Begin(\PhFetch{\Comb}){ \label{alg:pes_req_fetch}
		$\Phase \Leftarrow fetch$\;
		\ForEach{\Mol in \Comb}{
			dispatch \Fetch{\Mol}\;
		}
	}
	\Begin(\PhFetchResp{\RespMol}){ \label{alg:pes_req_fetch_resp}
		add \RespMol to \RArgs\;
		\If{all responses have arrived}{
			\Reaction{\RArgs}\;
		}
	}
	\Begin(\Abandon{\Comb}){
		$\Phase \Leftarrow none$\;
		\ForEach{\Mol in \Comb}{
			dispatch \GiveUp{\Mol}\;
		}
	}
\end{algorithm}

\end{minipage}
\hspace{0.1cm}
\begin{minipage}[t]{0.47\linewidth}
\vspace{0pt}
\begin{algorithm}[H]
	\caption{Pessimistic Protocol --- Holder.}
	\label{alg:pes_hol}
	\SetKwData{Msg}{message}
	\SetKwData{Sender}{sender}
	\SetKwData{Req}{req}
	\SetKwData{ReqI}{req\_i}
	\SetKwData{ReqJ}{req\_j}
	\SetKwData{Reacts}{no\_r}
	\SetKwData{Id}{id}
	\SetKwData{Mol}{molecule}
	\SetKwData{List}{list}
	\SetKwData{Locker}{locker}
	\SetKwData{RespOk}{RESP\_OK}
	\SetKwData{RespTaken}{RESP\_TAKEN}
	\SetKwData{RespRemoved}{RESP\_REMOVED}
	\SetKwData{Query}{QUERY}
	\SetKwData{GiveUp}{GIVE\_UP}
	\SetKwData{Commit}{COMMITMENT}
	\SetKwData{Fetch}{FETCH}
	\SetKwFunction{Sort}{SortRequesters}
	\Event(message received){ \label{alg:pes_hol_msg_recv}
		\If{\Msg = \GiveUp}{
			remove \Sender from \Mol.\List\;
		}
		\ElseIf{\Msg.\Mol does not exist}{
			reply with \RespRemoved\;
		}
		\ElseIf{\Msg = \Fetch}{
			clear \Mol.\List\;
			reply with \Mol\;
		}
		\ElseIf{\Mol has a commitment}{
			reply with \RespTaken\;
		}
		\ElseIf{\Msg = \Query}{
			add \Sender to \Mol.\List\;
			reply with \RespOk\; \label{alg:pes_hol_end_query}
		}
		\ElseIf{\Msg = \Commit}{ \label{alg:pes_hol_commit}
			\Sort{\Mol}\;
			\If{\Mol.\Locker = \Sender}{
				reply with \RespOk\;
			}
			\Else{
				reply with \RespTaken\;
			}
		}
	}
	\Begin(\Sort{\Mol}){ \label{alg:pes_hol_sort_begin}
		\ForEach{pair of requesters in \Mol.\List}{
			\If{\ReqJ.\Reacts $<$ \ReqI.\Reacts}{
				put \ReqJ before \ReqI\;
				\Cont\;
			}
			\If{\ReqJ.\Id $<$ \ReqI.\Id}{
				put \ReqJ before \ReqI\;
			}
		}
		\Mol.\Locker $\Leftarrow$ \Mol.\List(0)\; \label{alg:pes_hol_sort_end}
	}
\end{algorithm}
\end{minipage}
\end{table}

\subsection{Optimistic Sub-protocol}
\label{sec:protocols_opt}

\begin{figure}[!htpb]
\vspace{-15pt}
	\begin{minipage}[t]{0.46\linewidth}
		\centering
		\includegraphics[width=3.5cm]{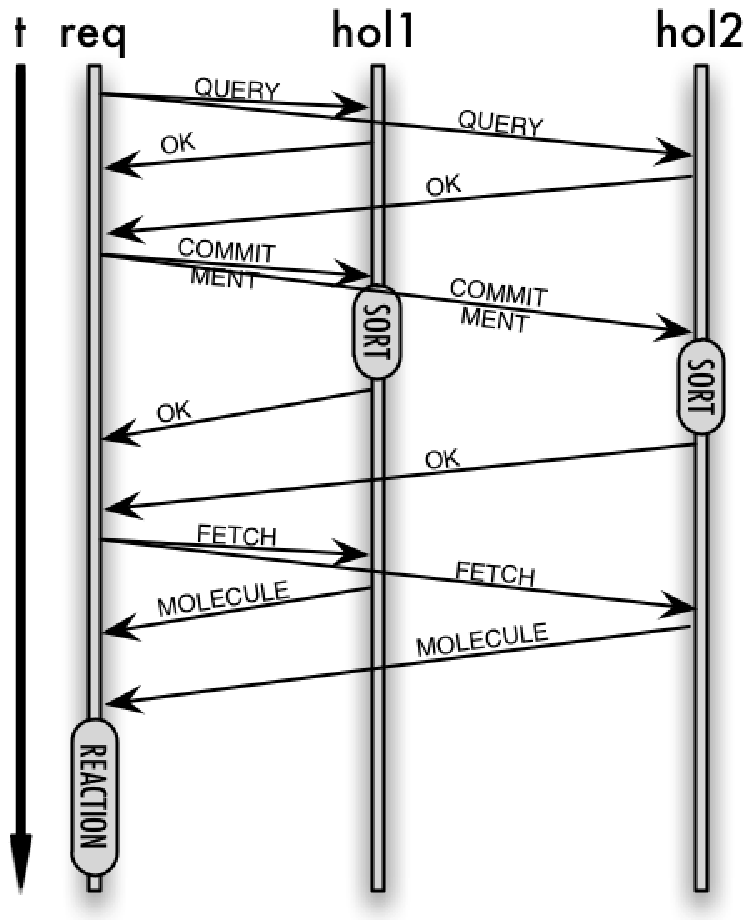}
		\caption{Pessimistic exchanges.}
		\label{fig:pessimist}
	\end{minipage}
	\hspace{0.15cm}
	\begin{minipage}[t]{0.46\linewidth}
		\centering
		\includegraphics[width=3.5cm]{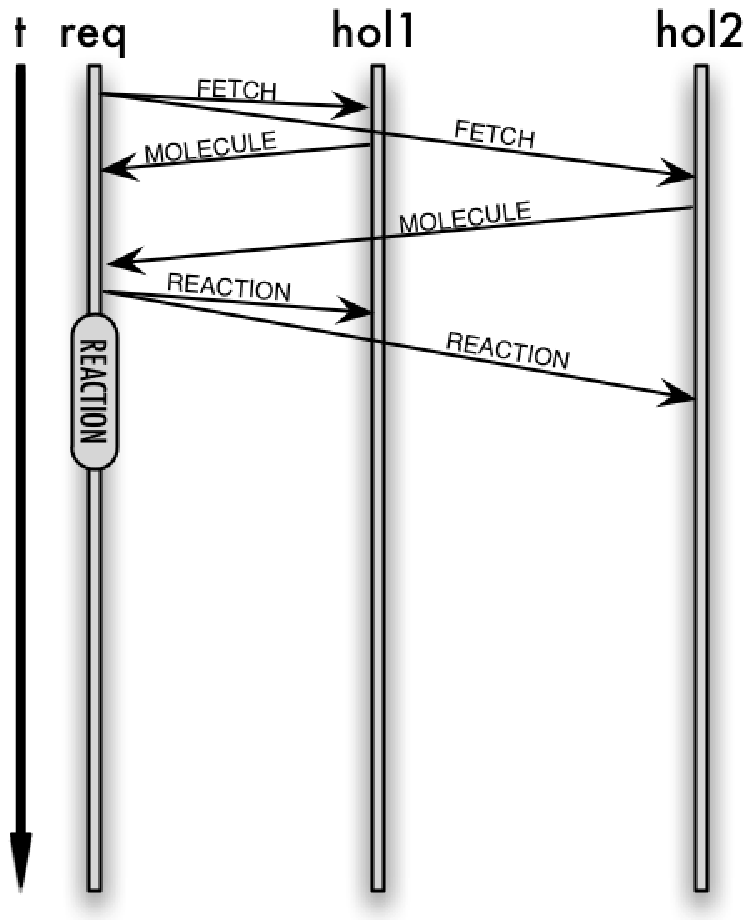}
		\caption{Optimistic exchanges.}
		\label{fig:optimist}
	\end{minipage}
\vspace{-25pt}
\end{figure}

When the possibility of multiple, concurrent reactions exists, the atomic fetch
procedure can be relaxed and simplified by adopting a more optimistic approach.
The optimistic sub-protocol requires only two stages --- the \emph{fetch} and the
\emph{notification} phases. Algorithm~\ref{alg:opt_req} describes the sub-protocol on
the requesters' side, while Algorithm~\ref{alg:opt_hol} describes it on the
holders' side. The time diagram of the process of obtaining molecules is
depicted in Figure~\ref{fig:optimist}.

%Requesters fetch molecules sequentially one by one, ordering them according to
%their hash identifiers. The molecule with the lowest id is fetched first and
%the one with the highest is fetched last.
Once a node has got information about suitable candidates, it immediately starts
the fetch phase (line~\ref{alg:opt_req_comb} in
Algorithm~\ref{alg:opt_req}). It dispatches \emph{FETCH} messages to the
appropriate holders. As with the pessimistic sub-protocol, the holder can respond
using the three previously described types of messages (\emph{RESP\_MOLECULE},
\emph{RESP\_TAKEN} and \emph{RESP\_REMOVED}) as shown in
Algorithm~\ref{alg:opt_hol}.
%\begin{enumerate}
%	\item \emph{RESP\_MOLECULE}: the requested molecule is available and is
%	transmitted to the requester,
%	\item \emph{RESP\_TAKEN}: the requested molecule has been fetched by another
%	node,
%	\item \emph{RESP\_REMOVED}: the requested molecule was used in a reaction and
%	no longer exists.
%\end{enumerate}
One holder that replied with a \emph{RESP\_MOLECULE} message, starts replying
with \emph{RESP\_TAKEN} messages to subsequent requests until the requester
either returns the molecule or notifies it a reaction took place.

%Depending on the holder's response, the requester takes different actions. If it
%received the molecule, it moves on to the next request. It does so until it
%either succeeds to grab all of the molecules needed for the reaction or until it
%receives a response other than \emph{RESP\_MOLECULE}. 
If the requester acquires all of the molecules, the reaction is subsequently
performed, and the requester sends out \emph{REACTION} messages to holders to notify
them the molecules are being consumed. This causes holders to reply with
\emph{RESP\_REMOVED} messages to subsequent requests from other requesters. In case
the requester received a \emph{RESP\_REMOVED} message, it aborts the reaction by
notifying holders with \emph{GIVE\_UP} messages, which allows holders to give
molecules to others.
%Finally, when a holder's reply is a \emph{RESP\_TAKEN}
%message, the requester waits a predefined amount of time and retries to fetch
%the molecule. If a maximal number of attempts has been exceeded, the requester
%aborts the reaction.

\paragraph{Conflict Resolution.} Given the fact that a node will most likely
execute the optimistic sub-protocol in a highly reactive stage, there is no need for
a strict conflict resolution policy. Instead, the node whose request first
reaches a holder obtains the desired molecule. However, the optimistic
sub-protocol does not ensure that a reaction will be performed in case of conflicts.

\begin{table}[!htpb]%[H]
%\vspace{-10pt}
\begin{minipage}[t]{0.48\linewidth}
\vspace{0pt}
\begin{algorithm}[H]
	\caption{Optimistic Protocol --- Requester.}
	\label{alg:opt_req}
	\SetKwData{Mol}{molecule}
	\SetKwData{Id}{id}
	\SetKwData{Holder}{holder}
	\SetKwData{Resp}{response}
	\SetKwData{RespMol}{resp\_mol}
	\SetKwData{Comb}{combination}
	\SetKwData{RArgs}{reaction\_args}
	\SetKwData{RespMol}{RESP\_MOLECULE}
	\SetKwFunction{Fetch}{FETCH}
	\SetKwFunction{WaitResp}{WaitForResp}
	\SetKwFunction{Abandon}{Abandon}
	\SetKwFunction{Reaction}{Reaction}
	\SetKwFunction{Notify}{NotifyHolders}
	\SetKwFunction{MsgReaction}{REACTION}
	\Event(combination found){ \label{alg:opt_req_comb}
		\ForEach{\Mol in \Comb}{
			dispatch \Fetch{\Mol}\;
		}
	}
	\Event(response received){
		\If{$\Resp \neq \RespMol$}{
			\Abandon{\Comb}\;
			\Return\;
		}
		add \Resp.\Mol to \RArgs\;
		\If{all responses have arrived}{
			\Notify{\Comb}\;
			\Reaction{\RArgs}\;
		}
	}
	\Begin(\Notify{\Comb}){
		\ForEach{\Mol in \Comb}{
			dispatch \MsgReaction{\Mol}\;
		}
	}
	\Begin(\Abandon{\Comb}){
		\ForEach{\Mol in \Comb}{
			dispatch \GiveUp{\Mol}\;
		}
	}
\end{algorithm}
\end{minipage}
\hspace{0.1cm}
\begin{minipage}[t]{0.5\linewidth}
\vspace{0pt}
\begin{algorithm}[H]
	\caption{Optimistic Algorithm --- Holder.}
	\label{alg:opt_hol}
	\SetKwData{Msg}{message}
	\SetKwData{Sender}{sender}
	\SetKwData{Mol}{molecule}
	\SetKwData{State}{state}
	\SetKwData{GiveUp}{GIVE\_UP}
	\SetKwData{RespOk}{RESP\_OK}
	\SetKwData{RespTaken}{RESP\_TAKEN}
	\SetKwData{RespRemoved}{RESP\_REMOVED}
	\SetKwData{RespMol}{RESP\_MOLECULE}
	\SetKwData{Reaction}{REACTION}
	\Event(message received){
		\If{\Msg = \GiveUp}{
			$\Mol.\State \Leftarrow free$\;
		}
		\ElseIf{\Msg = \Reaction}{
			remove \Mol\;
		}
		\ElseIf{\Msg.\Mol does not exist}{
			reply with \RespRemoved\;
		}
		\ElseIf{$\Mol.\State = taken$}{
			reply with \RespTaken\;
		}
		\Else{
			$\Mol.\State \Leftarrow taken$\;
			reply with \RespMol\;
		}
	}
\end{algorithm}
\end{minipage}
%\vspace{-10pt}
\end{table}

\subsection{Sub-protocol Mixing}
\label{sec:protocols_mix}

During its execution, a program typically can pass through two different 
stages. The first one is a highly reactive stage, which is characterised by a
high volume of possible concurrent reactions. In such a scenario, the use of the
pessimistic sub-protocol would lead to superfluous network traffic, since the
probability of a reaction's success is rather high. Thus, the optimistic
approach is enough to deal with concurrent accesses to molecules. The second
stage is the quiet stage, when there is a relatively small number of possible
reactions. Since this entails highly probable conflicts between nodes, the
pessimistic sub-protocol needs to be employed in order to ensure the advancement of
the system. Thus, the execution environment has to be able to adapt to changes
and \emph{switch} to the desired protocol accordingly. Moreover, these protocols
have got to be able to coexist in the same environment, as different nodes may
act according to different modalities at the same time.

Ideally, the execution environment should be perceived as a whole in which the
switch happens unanimously and simultaneously. Obviously, a global view of the
reaction potential cannot be maintained. Instead, each node independently
decides which protocol to employ for each reaction. The decision is first based
on a node's local success rate denoted $\sigma_{local}$, computed based on the
success history of the last queries the node issued. In order not to base the
decision only on its local observations, a node also keeps track of local
success rates of other nodes. Each time a node receives a request or a reply
message, the sender supplies it with its own current history-based success rate,
stored into another list of tunable size. We denote $\sigma$ the overall success
rate, computed as the weighted arithmetic mean of a node's local success rate
and the ones collected from other nodes. Finally, the decision as to which
protocol to employ depends on the rule a node wishes to execute. More
specifically, it is determined by the number of the rule's arguments, since the
more molecules the rule needs, the harder it is to assure they will be grabbed:
to grab $r$ molecules, a node employs the optimistic sub-protocol if and only if
$\sigma^r \geq s$, where $r$ is the number of arguments the chosen rule has and
$s$ is a predefined threshold value. If the inequality is not satisfied, the
node employs the pessimistic sub-protocol.

\subsubsection{Coexistence.}

Due to the locality of the switch between protocols, not all participants in the
system will perform it in the exact same moment, leading to possible
inconsistencies in the system, where some nodes try to grab the same molecules
using different protocols. In order to distinguish between \emph{optimistic} and
\emph{pessimistic} requests, each requester incorporates a \emph{request type}
field into the message being sent. Based on this field, the node holding the conflicting molecules
gives priority to nodes employing the more conservative, pessimistic
algorithm. Although this decision discourages \emph{optimistic} nodes and sets
them back temporarily, it ensures that, in the long run, \emph{eventually} a
node will be able to grab the molecules it needs, since pessimism is favoured
over optimism.

%optimistic algorithm. The reason for this decision is twofold. On one hand,
%giving priority to the pessimistic protocol would temporarily increase the
%number of messages in the system because of its three-phase nature. On the other
%hand, if a node is employing the optimistic protocol, it means it has perceived
%the system is in the highly reactive stage and should not be held back.

\subsection{Sketch of Proof for Correctness and Liveness}
\label{sec:protocols_proof}

The proposed protocol is a combination of the extensions of two existing
protocols presented in~\cite{3PC} and~\cite{Lampson79crashrecovery}. These two
protocols were initially introduced to guarantee resource transactions with only
one holder. In our context, a requester can ask for several molecules owned by
different holders. 

These protocols must guarantee two properties: i) \emph{correctness}: a molecule
is used in only one reaction (as we consider that every reaction consumes all of
the molecules entering it), and ii) \emph{liveness}: if a node sends a request
infinitely often, it will eventually succeed in capturing the molecules,
provided the requested molecules are still available.

%\begin{itemize}
%\item The correctness: a molecule is used in only one reaction (as
%  we consider than every reactions consume all the molecule getting in
%  the reaction).
%\item The liveness: if a node sends a request infinitely often, eventually
%  the request will sucesses, on condition that required molecules are
%  still available,
%\end{itemize}

\subsubsection{Correctness Proof.} 

Correctness is easy to prove because both protocols we rely on have been proved to be
correct independently. There are two cases of conflict between the two
protocols. When an optimistic request arrives before a pessimistic one, the
pessimistic request is aborted because the molecule has already been reserved by
the optimistic requester. On the other hand, if a pessimistic request arrives
first, the optimistic request is aborted in favour of the pessimistic one.

\subsubsection{Liveness Proof.}

To prove the liveness property, we show that: i) if no successful reaction
happens in the system, nodes eventually switch to the pessimistic protocol, ii)
if several pessimistic requesters are in conflict, at least one reaction is not
aborted, and iii) a node cannot see its reactions infinitely aborted.
%\begin{itemize}
%\item If no successful reaction happen in the system, nodes
%  eventually switch to pessimist protocol
%\item If several pessimist requests are in conflict at least one
%  reaction is not aborted
%\item A node can't have its reactions infinitely aborted.
%\end{itemize}

Initially, and hopefully most of the time, nodes use the optimistic sub-protocol
for their requests. In case of a conflict between two optimistic requesters,
both requests can easily be aborted. Consider the example where two concurrent
requesters try to capture two molecules, $A$ and $B$. If the first requester
succeeds in grabbing $A$ while the second captures $B$, then the two requests
will be aborted.

For the pessimistic protocol, we define a total order based on the number of
successfully completed reactions by a node and its id.
%\begin{itemize}
%\item the number of successful reaction
%\item the node's id
%\end{itemize}
In case of a conflict, all of the reactions might be aborted except for one --- the
reaction initiated by the node which comes first as per the total order. Because
the total order is based on the number of successful reactions, if a node, in
case of an abort, tries again infinitely to request molecules for its reaction,
eventually, if the requested molecules are still available, the reaction will
take place, given the fact that its position moves up the total ordering when
other nodes succeed in executing their reactions.

When a request of a node is aborted, the node decreases its value of $\sigma$
(see Section~\ref{sec:protocols_mix}). With each message sent, the node includes
the information about its local $\sigma$, and collects the values received with
each message. If there are many conflicts during a certain period of time, all
the more so if there is no successful reaction, the local $\sigma$ of all of the
nodes decreases. This effect leads to a situation where the value of the
computed $\sigma^r$ for all new reactions will be lower than the threshold
$s$, which will force the nodes to use the pessimistic protocol when
initiating new requests, which insures the system's liveness.

When presenting algorithms for atomic capture, it is common to study their
convergence times. However, any discussion about convergence when dealing with
the chemical programming model is not feasible, as convergence itself, and thus
the convergence time, is an application-specific property. However, the next
section presents an evaluation of the proposed algorithm, and sheds some light
on the subject.

\section{Evaluation}
\label{sec:sim}

Our protocol was simulated in order to better capture its performances. We
developed a Python-based discrete-time simulator, including a DHT layer
performing the random dissemination of a set of molecules over the nodes, on top
of which the layer containing the capture protocol itself was built. At this
layer, any message issued at step $t$ will be received and processed by the
destination node at time $t+1$. Moreover, each time a capture attempt either led
to a reaction, or an abortion, the node tries to fetch another set of $r$
randomly chosen molecules. Finally, on the top layer, a simple chemical
application was simulated. 

All presented experiments simulate a system of 250 nodes trying to execute a
chemical program containing a solution with 15 000 molecules and a
straightforward rule which simply consumes two molecules without producing new
ones. Such a simple program allows us to concentrate exclusively on evaluating
the capture protocol itself, without having to deal with application-specific
logic. In the same vein, reactions' duration are assumed negligible. Each
simulation was run $50$ times and the figures presented below show the values
obtained by averaging result data from these runs. Keep in mind that the final
steps of the executions shown in the figures represent, due to the effect of
averaging, worst-case scenarios obtained during simulation. Simulations were
limited to execute at most $500$ steps, as later steps are not relevant.

\begin{figure}[!htpb]
	\begin{minipage}[t]{0.47\linewidth}
		\centering
		\includegraphics[width=6.1cm]{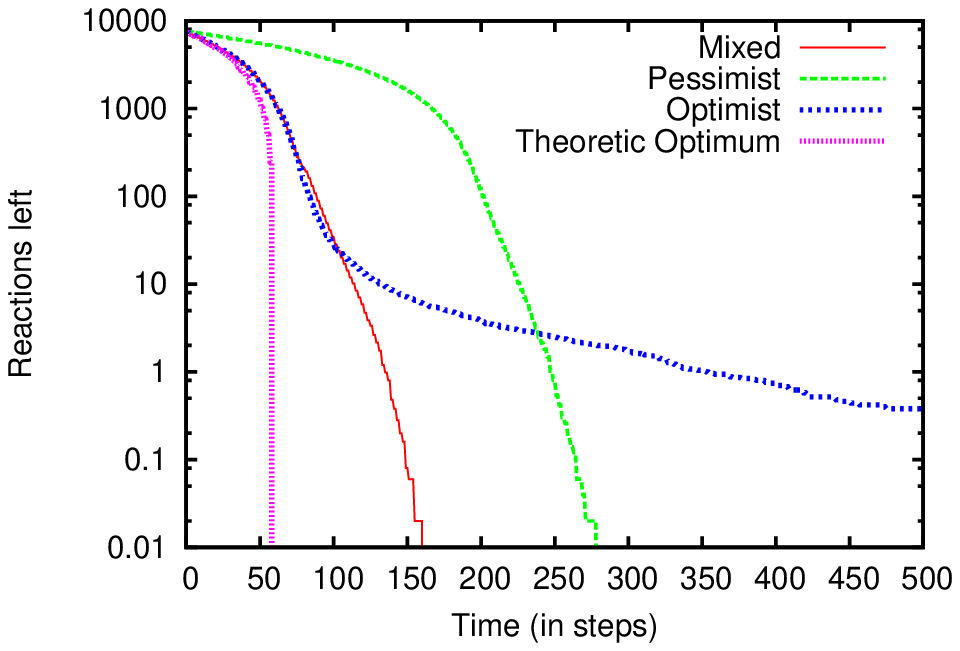}
		\caption{Performance comparison of the protocol's variants.}
		\label{fig:exp1}
	\end{minipage}
	\hspace{0.35cm}
	\begin{minipage}[t]{0.47\linewidth}
		\centering
		\includegraphics[width=6.1cm]{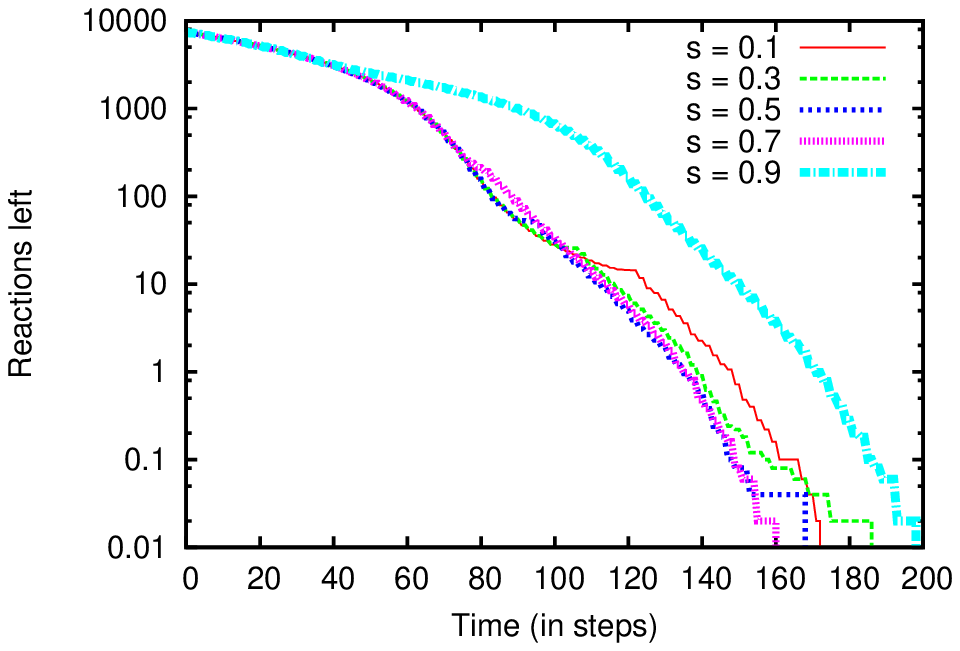}
		\caption{Execution time for different switch thresholds.}
		\label{fig:exp2}
	\end{minipage}
\end{figure}
\paragraph{Experiment 1.} Firstly we evaluate separately the performance
characteristics of both sub-protocols. Figure~\ref{fig:exp1} shows the
averaged number of reactions left to execute at each step, until inertia, using
only the optimistic mode, only the pessimistic mode, and the complete protocol
with switches between protocols (using $\sigma = 0.7$), respectively. Note that
a logarithmic scale is used. The figure shows that, using only the
\emph{optimistic} protocol, while we can see a strong decline in the number of
reactions left at the beginning of the computation, \emph{i.e.}, when a lot of
reactions are possible and that thus there are only few conflicts in the
requests, it gets harder for nodes to grab molecules when this number
declines. In fact, the system is not even able, for most of the executions, to
conclude the execution, as a few reactions left are never executed, always
generating conflict at fetch time. When the nodes are all \emph{pessimistic},
there is a steady, linear decrease in the number of reactions left, and the
system is able to reach the inertia in a reasonable amount of
time, thanks to the liveness ensured in this mode. For most
steps, the \emph{mixed} curve traces the exact same path as the
\emph{optimistic} one, which means that during this period the nodes employ the
optimistic sub-protocol. However, at the end, the system is able to quickly finish
the execution as an aftermath of switching to the pessimistic protocol. After
the switch, it diverges from the optimistic one to mimic the pessimistic curve,
exhibiting a benefit of a 42\% performance boost compared to the performance of
the pessimistic sub-protocol. Finally, the \emph{theoretic optimum} curve
represents the minimal amount of steps needed to complete the execution in a
centralised system. Comparing it to our protocol, we notice an increase of 166\%
in the number of steps needed to reach inertia. This is understandable, because
there is usually a coordinator in centralised systems with which conflict
situations can be circumvented, but it opens the door to serious defaults, such
as single-point-of-failure or bottleneck problems.

\begin{figure}[!htpb]
	\begin{minipage}[t]{0.47\linewidth}
		\centering
		\includegraphics[width=6.1cm]{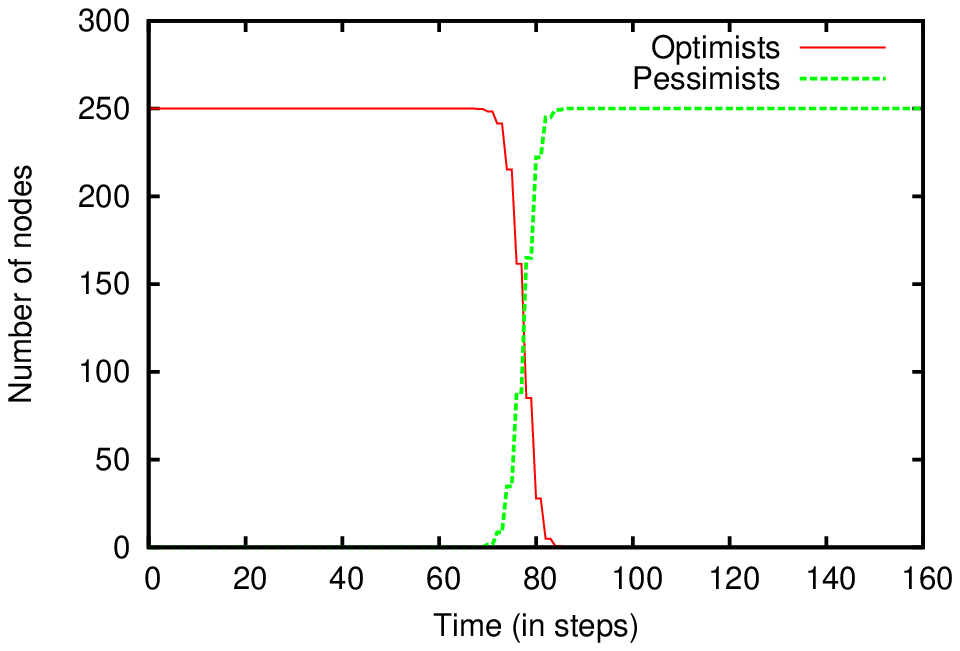}
		\caption{Number of nodes employing optimistic and pessimistic protocols per
		step.}
		\label{fig:exp3}
	\end{minipage}
	\hspace{0.35cm}
	\begin{minipage}[t]{0.47\linewidth}
		\centering
		\includegraphics[width=6.1cm]{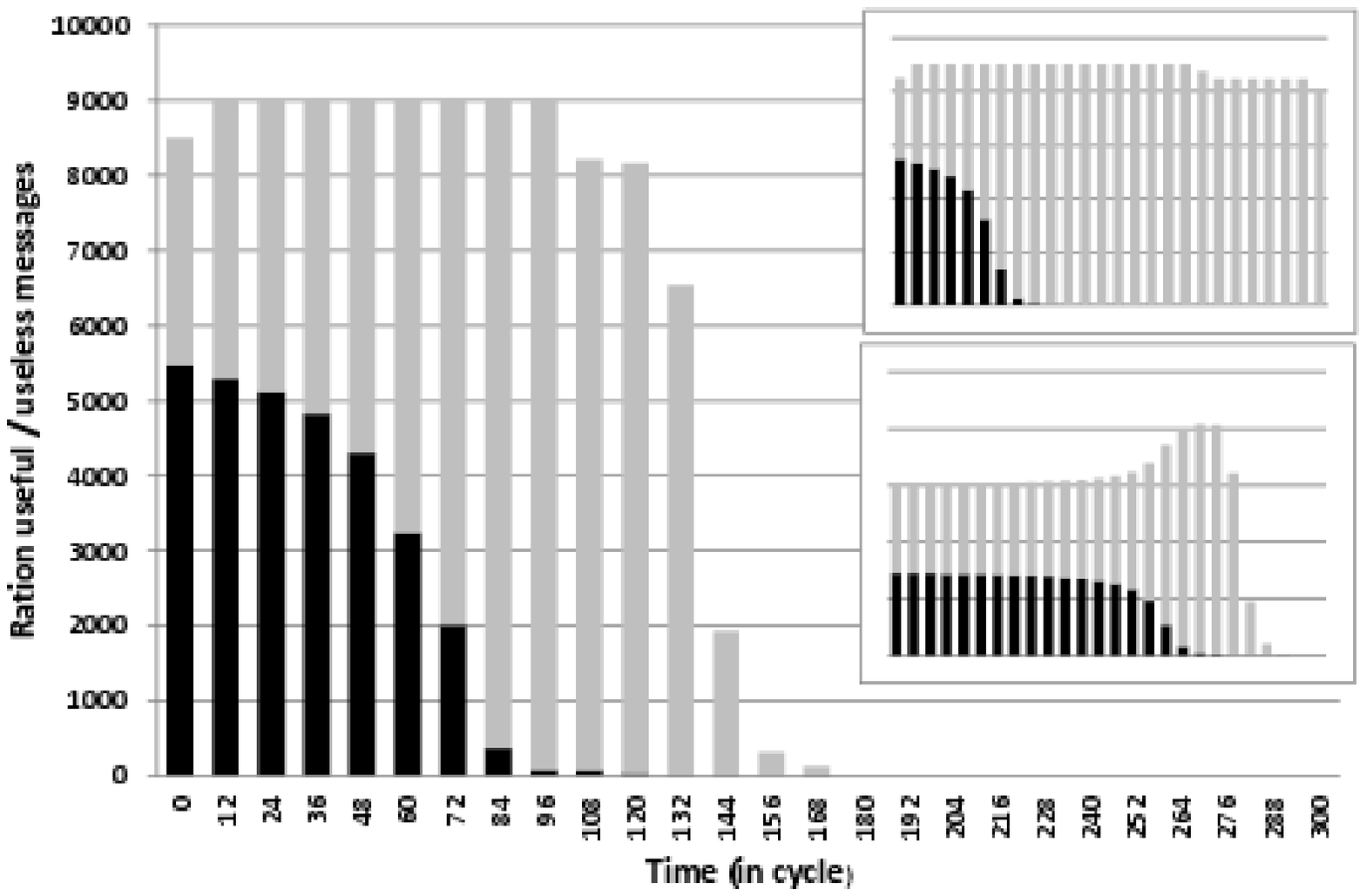}
		\caption{Number of messages sent per cycle.}
		\label{fig:exp4}
	\end{minipage}
\end{figure}

\paragraph{Experiment 2.} Next, we want to assess the impact of the switch
threshold $s$ on the overall performance of the system. Figure~\ref{fig:exp2}
depicts, in a logarithmic scale, the number of reactions left on each step for
different threshold values, varying from $0.1$ to $0.9$. As suspected, the
curves overlap during most steps, most nodes employing the
optimistic sub-protocol. The first curve to diverge is the one where the switch
threshold is set very high, to $s = 0.9$. Because the system depicted by that
curve did not fully exploit the optimistic sub-protocol, it is the last to finish
the execution. Although slightly, the other curves start diverging at different
moments, and, thus, complete the execution at different steps. Looking at
Figure~\ref{fig:exp2} brings us to the conclusion that, out of the five values
tested for the switch threshold, $s = 0.7$ yields the best performance results
in this particular scenario. Finding an overall optimal value for $s$ falls out
of the scope of this paper.

\paragraph{Experiment 3.} Here we examine the properties of the process of
switching from one protocol to the other. Figure~\ref{fig:exp3} shows that, at
the beginning of the execution, all of the nodes start off grabbing molecules by
using the optimistic sub-protocol. The switch happens about half way through the
execution. Around that time, \emph{optimistic} nodes can no longer efficiently
capture molecules, so they switch to the pessimistic sub-protocol. We observe
that, due to exchange of local $\sigma$ values, nodes in the system reach a
global consensus rather quickly --- for a system with 250 nodes, at most 15
steps are needed for all of the nodes to switch to the pessimistic protocol.%  We
% conclude that the system can respond quickly to changes in the system, even
% though there is no global view of the situation.

\paragraph{Experiment 4.} Finally, we investigate the communication costs
involved in the process. Figure~\ref{fig:exp4} depicts the number of messages
sent per cycle (where one cycle comprises $12$ simulation steps), classified into two
categories: useful messages (ones which led to a reaction, in black) and useless
messages (ones which did not induce a reaction, in grey). We note that the
protocol takes over the best properties of both of its sub-protocols. Firstly,
it takes over the elevated percentage of useful messages of the optimistic
sub-protocol. After the switch, the pessimistic protocol kicks in, bringing with
it a decrease in the total number of messages. When compared to the
communication costs of each of the sub-protocols separately (both depicted on
the right-hand side of Figure~\ref{fig:exp4}), we see that switching from one
protocol to the other reduces network traffic and improves scalability.

\section{Related Works}
\label{sec:related}

\vspace{-0.15cm}
The chemical paradigm was originally conceived for programs which need
to be executed on parallel machines. %In its early days, a lot of
%effort was put into parallel execution schemes for different types of
%existing parallel machines, notably
The pioneering work of Ban\^atre
\emph{et al.}~\cite{par_gamma_machine} provides two conceptual
approaches to the implementation problem, in both of which each processor of a parallel
machine holds a molecule and compares it with the molecules of all the
other processors.
%Two algorithms to achieve this are proposed: (a) a
%synchronous one, where a centralized controller triggers each
%comparison step, and (b) an asynchronous one, in which the molecules
%travel from processor to processor (each processor being connected to
%two neighbours on a vector), either until they react, or until they've
%returned to their starting point. The second algorithm has been
%implemented on top of an iPSC hypercube with 16 processors.
A slightly
different approach was proposed in the work of Linpeng \emph{et
al.}~\cite{par_gamma_massive}, where a program is executed
%on MasPar MP1, a massively parallel machine, using the fold-over operation. The
by placing
molecules on a strip, and then folding them over after each vertical
comparison.
%At each step, the elements in the upper segment of the
%strip are compared in parallel to those in the lower
%segment.
Recently, Lin \emph{et al.} developed a parser of GAMMA
programs for their execution on a cluster exploiting GPU computing
power~\cite{LKG10}. All works mentioned exhibit significant speed-up
properties, but the platforms experimented are rather restricted.
%they have not been conceived with long-running
%tasks in mind.
%and have been tested only on so-called \emph{reducer}
%rules, \emph{i.e.}, rules which produce fewer molecules than they
%consume, and thus uniformly reducing the complexity of the problem.

Mutual exclusion and resource allocation algorithms have been
studied extensively.
%subject of a wide number of works in the fields of parallel and
%distributed computing.
Nevertheless, most research focuses on sharing one
specific resource, or critical section, amongst many
processes~\cite{Sanders1987,Chandy1984}. A basic solution
for the \emph{k-out of-M} problem was given by
Raynal~\cite{Raynal1991}. This early work is a static permission-based
algorithm in which only the number of a predefined set of
resources varies from node to node. 
%every node keeps track of the number of resources
%currently in use by all the other nodes. To be able to use its $k_i$
%resources, the node $n_i$ needs to obtain permission from all of the
%other nodes, where requests are totally ordered using Lamport's
%timestamps. It is a static solution in which only the number of
%resources varies from node to node.
In addition, the solution supposes
a global knowledge of the system. On the other hand, an execution
environment for chemical programs is a dynamic system in which nodes
need to obtain different molecules, which can be thought of as
resources, at different times.

The three-phase commit protocol was originally proposed as a crash
recovery protocol for distributed database systems~\cite{3PC}. % The
% authors study the two-phase protocol and add to it a third, so-called
% \emph{prepare commit} phase.
%thanks to which they're able to obtain a
%system which is able to abort database transactions in any
%moment.
Although, in its essence similar to the three-phase commit
protocol, the goal of the optimistic sub-protocol proposed in this paper
is to secure the liveness of the system by ensuring that at least one
node will be able to complete its reaction.

\vspace{-0.25cm}
\section{Conclusion}
\label{sec:conclusion}

\vspace{-0.15cm}
While chemical metaphors are gaining attention in the modelling of
autonomous service coordination, the actual deployment of programs following the
chemical programming model over distributed platforms is a widely open
problem. In this paper, we have described a new protocol to capture several
molecules atomically in an evolving multiset of objects distributed on top of a
large-scale platform. By dynamically switching from one sub-protocol to
the other, our protocol fully exploits their good properties (the low communication overhead
and speed of the optimistic protocol, when the density of reactants is high,
and the liveness guarantee of the pessimistic protocol, when this density drops),
without suffering from their drawbacks. These features are illustrated by
simulation.

This protocol is part of an ambitious work which aims at building a
distributed autonomic platform providing all the features required to execute
chemical programs. This work is quite interesting in that it revisits classical
problems in distributed systems, but with the large scale requirements, as well
as the specificities of the chemical model, in mind. In this way, this paper
tackles the mutual exclusion: in our context the liveness property is a system
property while, more traditionally, liveness is a process' property. Among the
directions planned for this work, we will refine the execution model, to, for
instance, balance the load of reactions among the nodes of the platform. On the
practical side, we plan to use these algorithms to actually leverage the
expressiveness of the chemical paradigm for a workflow management system such as
defined in~\cite{FPT10}.

\vspace{-0.3cm}
\bibliographystyle{ieeetr}
\bibliography{main}

\end{document}